\documentstyle[preprint,eqsecnum,aps,epsf]{revtex}

\tightenlines
\begin{document}
\preprint{\vbox{
\hbox{INPP-UVA-00-03} 
\hbox{November 20, 2000} 
\hbox{hep-ph/0012295}
}}
\draft
\def\vp{{\bf p}}
\def\ko{K^0}
\def\kb{\bar{K^0}}
\def\al{\alpha}
\def\ab{\bar{\alpha}}
\def\be{\begin{equation}}
\def\en{\end{equation}}
\def\bea{\begin{eqnarray}}
\def\eea{\end{eqnarray}}
\def\non{\nonumber}
\def\la{\langle}
\def\ra{\rangle}
\def\epp{\epsilon^{\prime}}
\def\vep{\varepsilon}
\def\to{\rightarrow}
\def\up{\uparrow}
\def\dw {\downarrow}
\def\ms{\overline{\rm MS}}
\def\ums{{\mu}_{_{\overline{\rm MS}}}}
\def\u{\mu_{\rm fact}}

\def\arnps{{\sl Ann. Rev. Nucl. Part. Sci.}~}
\def\pr{{\sl Phys. Rev.}~}
\def\epj{{\sl Eur. Phys. J.}~}
\def\ijmp{{\sl Int. J. Mod. Phys.}~}
\def\jp{{\sl J. Phys.}~}
\def\mpl{{\sl Mod. Phys. Lett.}~}
\def\prp{{\sl Phys. Rep.}~}
\def\prl{{\sl Phys. Rev. Lett.}~}
\def\pl{{\sl Phys. Lett.}~}
\def\np{{\sl Nucl. Phys.}~}
\def\ppnp{{\sl Prog. Part. Nucl. Phys.}~}
\def\zp{{\sl Z. Phys.}~}

\title{Quark Orbital Angular Momentum in the Baryon\\}

\author{Xiaotong Song}

\address{Institute of Nuclear and Particle Physics\\ 
Department of Physics, University of Virginia, P.O.Box 400714\\
Charlottesville, Virginia 22904-4714\\}

\maketitle
\begin{abstract}
Analytical and numerical results, for the orbital and spin content 
carried by different quark flavors in the baryons, are given in the 
chiral quark model with symmetry breaking. The reduction of the quark 
spin, due to the spin dilution in the chiral splitting processes, is 
transferred into the orbital motion of quarks and antiquarks. The 
orbital angular momentum for each quark flavor in the proton as a 
function of the partition factor $\kappa$ and the chiral splitting 
probability $a$ is shown. The cancellation between the spin and 
orbital contributions in the spin sum rule and in the baryon magnetic 
moments is discussed. 
\end{abstract}

\bigskip
\bigskip
\bigskip

\pacs{12.39.Fe,~13.40.Em,~13.88.+e,~14.20.-c\\}

\newpage

\leftline{\bf I. Introduction}
\smallskip

Since the EMC data from a deep inelastic polarized muon-polarized proton 
scattering experiment was published \cite{emc89}, the study of spin became 
one of the most active areas of nuclear and particle physics. Through 
many experimental and theoretical investigations, spin physics has made 
significant progress in the past 
decade (see, for instance, most recent review papers \cite{hughes99}, 
\cite{lampe00}, and references therein). A basic question in the
study of nucleon spin is how the total angular momentum (spin) is 
distributed between the intrinsic spins of the quarks and gluons and 
their relative orbital motion. From quantum chromodynamics (QCD) \cite{ji}, 
the nucleon spin can be decomposed into the quark and gluon contributions
$$
{1\over 2}=\sum\limits_{q}<J_z>_{q+\bar q}+<J_z>_{G}={1\over
2}\Delta\Sigma+\sum\limits_{q}<L_z>_{q+\bar q}+<J_z>_{G},
\eqno (1)
$$
where $\Delta\Sigma/2=\sum\limits_q(\Delta q+\Delta\bar q)/2=
\sum\limits_q<s_z>_{q+\bar q}$ denotes the total contribution from the
intrinsic spin of quarks and antiquarks. Note that 
$\Delta q\equiv q_{\up}-q_{\dw}$, and 
$\Delta{\bar q}\equiv{\bar q}_{\up}-{\bar q}_{\dw}$, and $q_{\up,\dw}$ 
(${\bar q}_{\up,\dw}$) are quark (antiquark) {\it numbers} with spin 
parallel and antiparallel to the nucleon spin, or more precisely, quark
(antiquark) numbers of {\it positive} and {\it negative} helicities.
$\sum\limits_q<L_z>_{q+\bar q}$ denotes the total orbital angular momentum 
carried by quarks and antiquarks, and $<J_z>_{G}$ is the gluon angular 
momentum. Without loss of generality, in (1) the proton has been chosen 
to be {\it longitudinally polarized} in the $z$ direction; it has helicity 
of $+1/2$. The total angular momentum of quarks and antiquarks 
$<J_z>_{q+\bar q}$ has been decomposed into the spin and orbital parts in 
(1). 

Polarized deep-inelastic scattering (DIS) data \cite{smc-slac-hermes} 
indicate that the intrinsic quark spins only contribute about one third 
of the nucleon spin or even less. The smallness of ${1\over 2}\Delta\Sigma$ 
implies that the missing part should be contributed either by the quark 
orbital motion or the gluon angular momentum. There is no direct data on 
$\Delta G$ except for a preliminary restriction on $\Delta G(x)/G(x)$
given by experiment E581/704 \cite{e581}, and another indirect result 
$\Delta G\simeq 0.5-1.5$ at $Q^2\simeq 10$GeV$^2$ from the analysis of 
$Q^2$ dependence of $g_1(x,Q^2)$ \cite{smc97}. Several experiments for 
measuring $\Delta G$ \cite{gluon} have been suggested. For the orbital 
motion, it has been shown that $<J_z>_{q+\bar q}$ can be measured in 
the deep virtual Compton scattering (DVCS) \cite{ji-jws}, and one may 
then obtain the quark orbital angular momentum from the difference
$<J_z>_{q+\bar q}-<s_z>_{q+\bar q}$. Since the nucleon is a complex 
composite object of quarks and gluons, the configuration of these 
elementary constituents is determined by their mutual interactions 
governed by the underlying theory of strong interactions $-$ QCD. 
Study of different components in (1) and how they share the nucleon 
spin will give invaluable information on nonperturbative structure 
of the nucleon. 

Historically, when the quark model \cite{gellmann64} was invented 
in 1960's, all three quarks in the nucleon were assumed to be in 
S-states, so that $<L_z>_q=0$ and the nucleon spin arises entirely 
from the quark spin. On the other hand, in the simple parton model 
\cite{feynman69}, all quarks, antiquarks and gluons are moving in the 
same direction, i.e. parallel to the proton momentum, there is no 
transverse momentum for the partons and thus $<L_z>_{q+\bar q}=0$ 
and $<L_z>_G=0$. This picture cannot be $Q^2$ independent due to QCD 
evolution. In leading-log approximation, $\Delta\Sigma$ is $Q^2$
independent, but the gluon helicity $\Delta G$ increases with $Q^2$. 
This increase should be compensated by the decrease of the orbital 
angular momentum carried by partons (see for instance Ref. \cite{bms79} 
and later analysis \cite{rat87-sd89-qrrs90}). Recently, the leading-log 
evolution of $<L_z>_{q+\bar q}$ and $<L_z>_G$, and an interesting 
asymptotic partition rule were obtained in \cite{jth96}. The `$initial$'
value of the orbital angular momenta at the renormalization scale 
$\mu^2$ 
are determined by nonperturbative dynamics of the nucleon as a QCD 
bound state. Lattice QCD provides us with a nonperturbative tool to 
calculate the physical quantities of hadrons and leads to many 
interesting results \cite{lat1}. Meantime, many QCD inspired nucleon 
models have been developed to explain existing data and yield good 
physical insight into the nucleon. For instance, in the bag model 
\cite{ja-ji91,cmbag94}, $\sum\limits_q<s_z>_{q}\simeq 0.38$, and 
$\sum\limits_q<L_z>_{q}\simeq 0.12$, while in the Skyrme model
\cite{skyrme88}, $\Delta G=\Delta\Sigma=0$, and $<L_z>=1/2$, which
implies that the nucleon spin arises only from the orbital motion. 

Phenomenologically, long before the EMC experimental data \cite{emc89}
were published, using the Bjorken sum rule and low energy hyperon 
$\beta$-decay data, Sehgal \cite{sehgal} showed that nearly $40\%$ of 
the nucleon spin arises from the orbital motion of quarks, the remaining 
$60\%$ is attributed to the spin of quarks and antiquarks. Most recently 
Casu and Sehgal \cite{cs97} show that to fit the baryon magnetic moments 
and polarized DIS data, a large collective orbital angular momentum 
$<L_z>$, which contributes almost $80\%$ of nucleon spin, is needed. 
Hence the question remains of how much of the nucleon spin is coming 
from the quark orbital motion. 

Recently, the chiral quark model was successfully used to explain 
\cite{ehq92} both the smallness of $\Delta\Sigma$ and a flavor-asymmetry 
of the light-quark sea of the nucleon ($\bar d-\bar u\neq 0$). The 
model was improved by introducing U(1)-breaking \cite{cl1} and the 
kaonic suppression \cite{smw}. Several modified descriptions with 
both SU(3) and U(1)-breaking were developed in 
\cite{song9605,wsk,cl2,song9705,los}. The quark orbital motion in 
the chiral quark model was discussed in \cite{cl3} (SU(3) symmetry) 
and \cite{song9801332} (SU(3) breaking) respectively. In \cite{cl3}, 
authors found that due to significant cancellation between the 
contributions from the sea spin polarization and the orbital angular 
momentum, the naive quark model can yield a good account of the baryon 
magnetic moments. In \cite{song9801332}, an unified formalism for 
describing both spin and orbital angular momentum was suggested. 
However, an `{\it equal sharing}' assumption was used in 
\cite{song9801332} and needs to be improved. In this paper, the 
partition factor $\kappa$ [see section III (a)] is no longer 
restricted to be 1/3 and can be any value in the range of [0,1/2]. 
In section II, the basic assumptions and formalism are briefly 
reviewed. In section III, we give the formalism for calculating both 
spin and orbital angular momentum carried by quarks and antiquarks 
in the nucleon. Generalization to the octet and decuplet baryons is 
given in section IV. Numerical results are also given. The magnetic 
moments are discussed in section V. Different scenarios of the proton 
spin decomposition and a brief summary are given in section VI.
\bigskip

\leftline{\bf II. Chiral Quark Model}
\smallskip

The nonperturbative dynamics of quarks and gluons inside the nucleon 
is closely related to the QCD vacuum structure. At the energies 
lower than the nucleon mass ($\sim$ 1 GeV), the QCD vacuum has 
nonperturbative condensations of quarks \cite{gor68} and gluons 
\cite{svz79}. The nonzero vacuum expectation values imply that the 
$SU(3)\times SU(3)$ chiral symmetry is broken down to $SU(3)$ flavor 
symmetry. This phenomenon is manifested in the existence of eight 
Goldstone bosons $(\pi, K,\eta)$ and the existence of dynamical 
mass in the current quark propagator. Hence the bare massless 
(current) quarks become massive (constituent) quarks, which make a 
good basis to represent the QCD Hamiltonian at low energies. In the 
chiral quark model, the important degrees of freedom are quarks and 
Goldstone bosons, and the dominate interaction is the coupling among 
quarks and Goldstone bosons, while the gluon effect is expected to 
be rather small ($<J_z>_G\neq 0$ case will be discussed briefly in 
the later sections). In this description, the structure of the 
nucleon is determined by the valence quark structure of the nucleon 
and all possible quark-Goldstone boson fluctuations. The sea of 
quarks and antiquarks in this model is generated from the Goldstone 
bosons and not from gluons. 

Following the notations used in \cite{song9705} and \cite{song9801332}, 
the effective Lagrangian is 
$${\it L}_I=g_8{\bar q}\pmatrix{{G}_u^0
& {\pi}^+ & {\sqrt\epsilon}K^+\cr 
{\pi}^-& {G}_d^0
& {\sqrt\epsilon}K^0\cr
{\sqrt\epsilon}K^-& {\sqrt\epsilon}{\bar K}^0
&{G}_s^0 
\cr }q, 
\eqno (2a)$$
where ${G}_{u(d)}^0$ and ${G}_{s}^0$ are defined as
$${G}_{u(d)}^0=+(-){{\pi^0}\over{\sqrt 2}}+
{\sqrt{\epsilon_{\eta}}}{{\eta^0}\over{\sqrt 6}}+
{\zeta'}{{\eta'^0}\over{\sqrt 3}},
~~~{G}_s^0=-{\sqrt{\epsilon_{\eta}}}{{\eta^0}\over{\sqrt 6}}+
{\zeta'}{{\eta'^0}\over{\sqrt 3}},
\eqno (2b)$$
and the breaking effects are explicitly included. We use $a\equiv|g_8|^2$ 
to denote the transition probability of chiral fluctuation or splitting 
$u(d)\to d(u)+\pi^{+(-)}$, then $\epsilon a$ denotes the probability 
of $u(d)\to s+K^{-(0)}$. Similar definitions are used for 
$\epsilon_\eta a$ and $\zeta'^2a$. If the breaking parameter is
dominated by the {\it mass suppression effect}, one reasonably expects
$0\leq\zeta'^2a<\epsilon_{\eta}a\simeq\epsilon a\leq a$, then we have 
$0\leq\zeta'^2\leq 1$, $0\leq\epsilon_{\eta}\leq 1$, and
$0\leq\epsilon\leq 1$. We note that the parameter $\zeta'$ must be 
{\it negative} which is required from the experimental data (see 
discussion in \cite{song9605}). In our numerical calculation, we use 
the approximation $\epsilon_\eta\simeq\epsilon$ given in \cite{song9705},
so that only $three$ parameters remain. However, our analytical results
and conclusions do not depend on this approximation. We also note that 
in our formalism, only the {\it integrated} quark spin and flavor content 
are discussed. 

The basic {\it assumptions} of the chiral quark model are: (i) The 
quark flavor, spin and orbital content of the nucleon are determined 
by its valence quark structure and all possible chiral fluctuations, 
and probabilities of these fluctuations depend on the interaction 
Lagrangian (2a-b). (ii) The coupling between the quark and Goldstone 
boson (GB) is rather weak, one can treat the fluctuation $q\to q'+{\rm
GB}$ as a small perturbation ($a\simeq 0.10-0.15$) and contributions 
from the higher order fluctuations can be neglected. (iii) The quark 
spin-flip interaction dominates the splitting process $q\to q'+{\rm GB}$.
This can be related to the picture given by the instanton model \cite{ins} 
and the spin-nonflip interaction is suppressed.

Based upon these assumptions, the quark {\it flips} its spin and 
changes (or maintains) its flavor by emitting a charged (or neutral) 
Goldstone boson. The light quark sea asymmetry $\bar u<\bar d$ is 
attributed to the existing {\it flavor asymmetry} of the valence quark 
numbers (two valence $u$-quarks and one valence $d$-quark) in the 
proton. On the other hand, the quark spin reduction is due to the 
{\it spin dilution} in the chiral splitting processes. Furthermore,
the quark spin component changes one unit of angular momentum,
$(s_z)_f-(s_z)_i=+1$ or $-1$, due to spin-flip in the fluctuation 
with GB emission. The angular momentum conservation requires the 
{\it same} amount change of the orbital angular momentum but with 
{\it opposite sign}, i.e. $(L_z)_f-(L_z)_i=-1$ or $+1$. This {\it 
induced} orbital motion is distributed among the quarks and antiquarks, 
and compensates the spin reduction in the chiral splitting. This is 
the starting point to calculate the orbital angular momentum carried 
by quarks and antiquarks in the chiral quark model. 

For spin-up or spin-down valence $u$, $d$, and $s$ quarks, up to the 
first order fluctuation, the allowed chiral processes are
$$u_{\up,(\dw)}\to d_{\dw,(\up)}+\pi^+,~~
u_{\up,(\dw)}\to s_{\dw,(\up)}+K^+,~~
u_{\up,(\dw)}\to u_{\dw,(\up)}+{G}_u^0,~~
u_{\up,(\dw)}\to u_{\up,(\dw)}.
\eqno (3a)$$
$$d_{\up,(\dw)}\to u_{\dw,(\up)}+\pi^-,~~d_{\up,(\dw)}\to 
s_{\dw,(\up)}+K^{\rm 0},~~
d_{\up,(\dw)}\to d_{\dw,(\up)}+{G}_d^0,~~
d_{\up,(\dw)}\to d_{\up,(\dw)},
\eqno (3b)$$
$$s_{\up,(\dw)}\to u_{\dw,(\up)}+K^-,~~
s_{\up,(\dw)}\to d_{\dw,(\up)}+{\bar K}^{\rm 0},~~
s_{\up,(\dw)}\to s_{\dw,(\up)}+{G}_s^{\rm 0},~~
s_{\up,(\dw)}\to s_{\up,(\dw)}.
\eqno (3c)$$
We note that the quark spin flips in the chiral splitting processes 
$q_{\up,(\dw)}\to q_{\dw,(\up)}$+GB, i.e. the first three processes 
in each of (3a), (3b), and (3c), but not for the last one. In the
zeroth approximation, the SU(3)$\otimes$SU(2) proton wave function
gives 
$$n^{(0)}_p(u_{\up})={5\over 3}~,~~~n^{(0)}_p(u_{\dw})={1\over 3}~,~~~
n^{(0)}_p(d_{\up})={1\over 3}~,~~~n^{(0)}_p(d_{\dw})={2\over 3}~.
\eqno (4)$$
the spin-up and spin-down quark (or antiquark) content, up to first
order fluctuation, can be written as
$$n_p(q'_{\up,\dw}, {\rm or}\ {\bar q'}_{\up,\dw}) 
=\sum\limits_{q=u,d}\sum\limits_{h=\up,\dw}
n^{(0)}_p(q_h)P_{q_h}(q'_{\up,\dw}, {\rm or}\ {\bar q'}_{\up,\dw}),
\eqno (5)$$
where $P_{q_{\up,\dw}}(q'_{\up,\dw})$ and $P_{q_{\up,\dw}}({\bar
q}'_{\up,\dw})$ are probabilities of finding a quark $q'_{\up,\dw}$
or an antiquark $\bar q'_{\up,\dw}$ arise from all chiral fluctuations 
of a valence quark $q_{\up,\dw}$. 
$P_{q_{\up,\dw}}(q'_{\up,\dw})$ and $P_{q_{\up,\dw}}({\bar q}'_{\up,\dw})$ 
can be obtained from the effective Lagrangian (2) and listed in Table I,
where only $P_{q_{\up}}(q'_{\up,\dw})$ and $P_{q_{\up}}({\bar
q}'_{\up,\dw})$ are shown. Those arise from $q_{\dw}$ can be obtained by
using the relations, $P_{q_{\dw}}(q'_{\up,\dw})=P_{q_{\up}}(q'_{\dw,\up})$
and $P_{q_{\dw}}({\bar q}'_{\up,\dw})=P_{q_{\up}}({\bar q}'_{\dw,\up})$.
The notations given in Table I are defined as 
$$A\equiv 1-\zeta'+{{1-{\sqrt\epsilon_{\eta}}}\over 2},
\qquad  B\equiv \zeta'-{\sqrt\epsilon_{\eta}},\qquad
C\equiv \zeta'+2{\sqrt\epsilon_{\eta}}.
\eqno (6a)$$
and 
$$f\equiv{1\over 2}+{{\epsilon_{\eta}}\over 6}+{{\zeta'^2}\over 3},~~~
f_s\equiv{{2\epsilon_{\eta}}\over 3}+{{\zeta'^2}\over 3},
\eqno (6b)$$
The special combinations $A$, $B$, and $C$ stem from the quark and
antiquark contents in the octet and singlet neutral bosons ${G}_{u(d)}^0$ 
and ${G}_{s}^0$ [see (2b)] appeared in the effective chiral Lagrangian 
(2a), while $f$ and $f_s$ stand for the probabilities of the chiral 
splittings $u_{\up}(d_{\up})\to u_{\dw}(d_{\dw})+{G}_{u(d)}^0$ and
$s_{\up}\to s_{\dw}+{G}_s^0$ respectively. Although there is no valence 
$s$ quark in the proton and neutron, there are one or two valence $s$ 
quarks in $\Sigma$ or $\Xi$, or other strange decuplet baryons, and even
three valence $s$ quarks in the $\Omega^-$. Hence for the purpose of later 
use we also give the probabilities arise from a valence $s$-quark splitting. 
In general, the suppression effects may be different for different baryons, 
hence the probabilities $P_{q_{\up,\dw}}(q'_{\up,\dw})$ and 
$P_{q_{\up,\dw}}({\bar q}'_{\up,\dw})$ may $vary$ with the baryons.
But we will assume that they are $universal$ for all baryons. 

Using (4), (5) and probabilities listed in Table I, the spin-up and 
spin-down quark and antiquark content, and the spin average and spin 
weighted quark and antiquark content in the proton were obtained in 
\cite{song9605,song9705} and are now collected in Table II. For the 
purpose of later discussion, we write down the formula for the 
spin-weighted quark content
$$(\Delta q')^{B}=\sum\limits_q [n^{(0)}_B(q_{\up})-n^{(0)}_B(q_{\dw})]
[P_{q_{\up}}(q'_{\up})-P_{q_{\up}}(q'_{\dw})].
\eqno (7a)$$
Note that the spin-weighted antiquark content is zero
$$(\Delta\bar q')^B=0.
\eqno (7b)$$
Hence one has $(\Delta q)_{sea}\neq \Delta\bar q$ in the chiral quark 
model. This is different from those models, in which the sea quark and 
antiquark with the same flavor are created as a pair from the gluon 
(see discussion on Eq.(3.12) in \cite{smw}). The quark spin contents in
the proton are 
$$\Delta u^p={4\over 5}\Delta_3-a,~~\Delta d^p=-{1\over 5}\Delta_3-a,~~
\Delta s^p=-\epsilon a,
\eqno (7c)$$
where $\Delta_3\equiv\Delta u^p-\Delta d^p={5\over 3}[1-a(\epsilon+2f)]$.
The total quark spin content in the proton is
$${1\over 2}\Delta\Sigma^p={1\over 2}(\Delta u^p+\Delta d^p+\Delta
s^p)={1\over 2}-(1+\epsilon+f)a\equiv {1\over 2}-\xi_1a, 
\eqno (7d)$$
where the notation $\xi_1\equiv 1+\epsilon+f$ is introduced. The total
{\it loss} of quark spin $(1+\epsilon+f)a$ appeared in Eq.(7d) is due to 
the fact that there are {three} splitting processes with quark spin-flip
[see the first three processes in (3a) and (3b)], probabilities of 
these spin-flip splittings are $a$, $\epsilon a$, and $fa$ respectively. 
 
\bigskip

\leftline{\bf III. Quark Orbital Motion}
\smallskip

\leftline{\quad (a)~\bf Quark orbital angular momentum in the nucleon}
 
The quark orbital angular momentum can be discussed in a similar way. 
For instance, for a spin-up valence $u$-quark, only first three 
processes in (3a), i.e. quark fluctuations with {\rm GB} emission, 
can induce a change of the orbital angular momentum. The last process 
in (3a), $u_{\up}\to u_{\up}$ means no chiral fluctuation and it makes 
no contribution to the orbital motion and will be disregarded. The 
orbital angular momentum produced in the splitting $q_{\up}\to q'_{\dw}
+{\rm GB}$ is shared by the recoil quark ($q'$) and the Goldstone boson
(GB). If we define the fraction of the orbital angular momentum shared
by the recoil quark is $1-2\kappa$, then the orbital angular momentum 
shared by the (GB) is $2\kappa$ which, we assume, equally shared by the
quark and antiquark in the Goldstone boson. We call $\kappa$ the {\it
partition factor}, which can take any value in the range [0,1/2]. For
$\kappa=1/3$, the three particles - the quark and antiquark in the (GB) 
and the recoil quark - equally share the induced orbital angular momentum.
This is the `{\it equal sharing}' case discussed in
\cite{song9801332}.
 
We define $<L_z>_{q'/q_{\up}}$ ($<L_z>_{{\bar q'}/q_{\up}}$) as the 
orbital angular momentum, carried by the quark $q'$ (antiquark $\bar q'$), 
arises from all fluctuations of a valence spin-up quark except for 
no-splitting case. Considering the quark spin component changes one unit 
of angular momentum in each splitting and using Table I, all 
$<L_z>_{q'/q_{\up}}$ and $<L_z>_{\bar q'/q_{\up}}$ for $q=u,d,s$ are
obtained and listed in Table III.

The orbital angular momentum produced from a spin-down valence quark 
splitting is the same as that from a spin-up valence quark splitting, 
but with {\it opposite sign}
$$ <L_z>_{q'/q_{\dw}}=-<L_z>_{q'/q_{\up}},~~ 
<L_z>_{{\bar q'}/q_{\dw}}=-<L_z>_{{\bar q'}/q_{\up}},
\eqno (8)$$
where both $q'_{\up}$ and $q'_{\dw}$ are included
in $<L_z>_{q'/q_{\up,\dw}}$ (the same is true for $<L_z>_{{\bar
q'}/q_{\up,\dw}}$).

Having obtained the orbital angular momenta carried by different 
quark flavors produced from the spin-up and spin-down valence quark
fluctuations, it is easy to write down the total orbital angular 
momentum carried by a specific quark flavor, for instance, $u$-quark, 
in the proton
$$<L_z>_{u}^p=\sum\limits_{q=u,d}
[n^{(0)}_p(q_{\up})-n^{(0)}_p(q_{\dw})]<L_z>_{u/q_{\up}},
\eqno (9)$$
where $\sum$ summed over {\it valence} $u$- and $d$-{\it quarks} in
the proton. $n^{(0)}_p(q_{\up})$ and $n^{(0)}_p(q_{\dw})$ are given 
in (4). Similarly, one obtains the $<L_z>_{d}^p$, $<L_z>_{s}^p$, and 
those for antiquarks. 

The analytical results are
$$\sum\limits_{q}<L_z>_{q}^p
=(1-\kappa)\xi_1a,
\eqno (10a)$$
$$\sum\limits_{\bar q}<L_z>_{\bar q}^p
=\kappa\xi_1a,
\eqno (10b)$$
$$\sum\limits_{q}<L_z>_{q+\bar q}^p\equiv
\sum\limits_{q}<L_z>_{q}^p+\sum\limits_{\bar q}<L_z>_{\bar q}^p=\xi_1a.
\eqno (10c)$$
Several remarks are in order: 
\begin{itemize}
\item{The amount $\xi_1a$ in (10c) is {\it exactly the same as} the total
spin reduction in (7d). The sum of (10c) and (7d) gives
$$
\sum\limits_q<J_z>_{q+\bar q}^p=
\sum\limits_{q}[<s_z>_{q+\bar
q}^p+<L_z>_{q+\bar q}^p]={1\over 2}.
\eqno (10d)$$
Therefore, in the chiral fluctuations, the missing part of the quark 
spin is {\it transferred} into the orbital motion of quarks and antiquarks. 
The amount of quark spin reduction $(1+\epsilon+f)a$ in (7d) is canceled
by the equal amount increase of the quark orbital angular momentum in
(10c), and the total angular momentum of nucleon is unchanged.} 

\item{The orbital angular momentum of quarks or antiquarks may depend on
the partition factor $\kappa$ [see(10a-b)], but the total orbital angular
momentum (10c) is {\it independent of} $\kappa$. This is because 
the total transfer of the angular momentum from `spin' into `orbital'
should be entirely determined by the chiral splitting not the partition
factor.}

\item{From (10b), the orbital angular momentum carried by antiquarks
$\sum\limits_{\bar q}<L_z>_{\bar q}$ increases with $\kappa$. This is
because a larger $\kappa$ means a larger portion of induced orbital 
angular momentum is transfered to the Goldstone boson, and thus to 
the antiquark. On the contrary, $\sum\limits_q<L_z>_{q}$ decreases 
with $\kappa$ [see (10a)].} 

\item{Although the orbital angular momentum carried by quarks ( 
antiquarks) $\sum\limits_q<L_z>_{q}^p$ ($\sum\limits_{\bar q}
<L_z>_{\bar q}^p$) depends on both $\kappa$ and the chiral parameters,
the ratio of $\sum\limits_q<L_z>_{q}^p$ and $\sum\limits_q<L_z>_{\bar
q}^p$ depends only on the partition factor $\kappa$ 
$$\sum\limits_q<L_z>_{q}^p/\sum\limits_q<L_z>_{\bar q}^p=(1-\kappa)/\kappa
\eqno(11a)$$ 
For $\kappa=1/3$ (equal sharing), this ratio is 2:1. This is originated
from the mechanism of the chiral fluctuation, there are {\it two} quarks
and {\it one} antiquark in the final state.}

\item{Contrary to (11a), the ratio of quark spin and orbital content is
$$  \sum\limits_q<s_z>_{q+\bar q}^p/\sum\limits_q<L_z>_{q+\bar q}^p=
{1\over {2\xi_1a}}-1,
\eqno (11b)$$
which is {\it independent of} $\kappa$, but depends on the chiral
parameters.}
\end{itemize}

The discussion can be easily extended to other baryons. Note that 
different baryons have different valence quark structure and thus 
different $n^{(0)}_B(q_{\up})$ and $n^{(0)}_B(q_{\dw})$. For the neutron,
explicit calculations show that $<L_z>_{u,\bar u}^n=<L_z>_{d,\bar d}^p$,
$<L_z>_{d,\bar d}^n=<L_z>_{u,\bar u}^p$, and
$<L_z>_{s,\bar s}^n=<L_z>_{s,\bar s}^p$. Using these relations, one can 
obtain the orbital angular momenta carried by quarks and antiquarks in 
the neutron. We have similar relations for $\Delta q$ from the 
isospin symmetry, hence results (7d), (10a-10d), (11a), (11b), and related
conclusions hold for the neutron as well. Extension to other octet and 
decuplet baryons will be given in section IV.
\smallskip

\leftline{\quad (b)~\bf Numerical results}

To determine model parameters, we use similar approach given in
\cite{song9705}, where the chiral quark model with only $three$
parameters gave a good description to most existing spin and flavor 
observables. The chiral parameters $a$, $\epsilon\simeq\epsilon_\eta$, 
and $\zeta'$ are determined by three inputs, $\Delta u-\Delta d=1.26$
\cite{pdg00}, $\Delta u+\Delta d-2\Delta s=0.60$ \cite{pdg00}, and 
$\bar d-\bar u=0.143$ \cite{nmc}. The good agreement between the model 
prediction and spin-flavor data can be seen from Table IV. 
The three-parameter set is: a=0.145,
$\epsilon=0.46$, and $\zeta'^2=0.10$. It gives
$$\xi_1\equiv 1+\epsilon+f=2.07.
\eqno (12)$$ 
Numerical results of the spin and orbital angular momentum shared by each
quark flavor are listed in Table V. Several comments are in order: 
\begin{itemize}
\item{From (10c) and (12), one obtains
$$\sum\limits_q<L_z>_{q+\bar q}^p\simeq 0.30,
\eqno (13)$$ 
i.e., nearly $60\%$ of the 
proton spin is coming from the orbital motion of quarks and antiquarks, 
and $40\%$ is contributed by the quark and antiquark spins. Comparison 
of our result with other models is given in Table VI and Fig. 1.}
\item{Using the parameter set given above, 
we have $\epsilon a\simeq\epsilon_\eta a\simeq 0.067$ and
$\zeta'^2a\simeq 0.015$. It implies that probabilities of chiral 
splittings $q\to q'\pi$, $q\to q'K (\eta)$, and $q\to q'\eta'$ 
are $14-15\%$, $6-7\%$, and $1-2\%$ respectively.}
\item{We plot the orbital angular momenta carried by quarks and 
antiquarks in the proton as functions of $\kappa$ in Fig.2. It shows 
that $<L_z>^p_s=<L_z>^p_{\bar s}$ at $\kappa=1/3$. This is because 
there is no valence $s$ quark in the nucleon. Hence there is no recoil 
$s$ quark in the chiral splitting, all $s$ and $\bar s$ can only come 
from the GB emission.}
\item{Comparing our choice of parameters with two extreme cases, 
$$\xi_1\equiv 1+\epsilon+f=3.0,~~~~~~~ {\rm for}~~{\rm U(3)-symmetry}~
(\epsilon=\epsilon_{\eta}=\zeta'^2=1)
\eqno (14a)$$
$$\xi_1\equiv 1+\epsilon+f=1.5,~~~~~~~ {\rm for}~~{\rm extreme~ breaking}~
(\epsilon=\epsilon_{\eta}=\zeta'^2=0)
\eqno (14b)$$
we see that the value $\xi_1$ given in (12) is just between
those given in (14a) and (14b).} 
\item{As indicated in Eq.(10d), the total quark spin reduction is 
canceled by the equal increase of the orbital angular momentum. However, 
the {\it exact} cancellation does {\it not} apply to each quark flavor. 
Using the superscript (0) denotes the quantity in NQM, one has 
$\Delta u^{p(0)}=4/3$, $\Delta d^{p(0)}=-1/3$, and $\Delta s^{p(0)}=0$.
From Table V, taking $\kappa=1/3$, we have in the chiral quark model 
$${1\over 2}\Delta u^p+<L_z>_{u+\bar u}^p=0.558~<~{1\over 2}\Delta
u^{p(0)}=0.667,
\eqno(15a)$$
$${1\over 2}\Delta d^p+<L_z>_{d+\bar d}^p=-0.080~>~{1\over 2}\Delta
d^{(0)p}=-0.167,
\eqno(15b)$$
$${1\over 2}\Delta s^p+<L_z>_{s+\bar s}^p=0.022~>~{1\over 2}\Delta
s^{(0)p}=0.
\eqno(15c)$$
We see that for the $u$-flavor, the orbital contribution is not 
big enough to compensate the quark spin reduction, but for both 
$d$-flavor and $s$-flavor the orbital contribution is too big and
the spin reduction is over-compensated. However, taking the sum, 
one has
$$\sum\limits_{q}[{1\over 2}\Delta q^p+<L_z>_{q+\bar q}^p]=
{1\over 2}=\sum\limits_{q}{1\over 2}\Delta q^{p(0)}.
\eqno (15d)$$}
\item{If we define $\Delta q_{sea}\equiv\Delta q-\Delta q_{val}$ and
identify $\Delta q_{val}=\Delta q^{(0)}$ which are determined from the
NQM, then 
$${1\over 2}\Delta u_{sea}^p+<L_z>_{u+\bar u}^p=-0.109,
\eqno (15e)$$
$${1\over 2}\Delta d_{sea}^p+<L_z>_{d+\bar d}^p=~0.087,
\eqno (15f)$$
$${1\over 2}\Delta s_{sea}^p+<L_z>_{s+\bar s}^p=~0.022. 
\eqno (15g)$$
They are {\it nonzero}. Hence the exact cancellation {\it does not 
occur} for each quark flavor, though it {\it does} for the sum of all 
quark flavors,  i.e. 
$$\sum\limits_{q}[{1\over 2}\Delta q_{sea}^p+<L_z>_{q+\bar q}^p]=0.
\eqno (15h)$$}
\end{itemize}
\bigskip

\leftline{\bf IV. Extension to other Baryons}
\smallskip

\leftline{\quad (a)~\bf Spin content in octet baryons}

We take $\Sigma^+(uus)$ as an example, other octet baryons can be 
discussed in a similar manner. The valence quark structure of 
$\Sigma^+$ is the same as the proton with the replacement
$d\to s$. Hence one has
$$n^{(0)}_{\Sigma^+}(u_{\up})={5\over 3}~,~~~
n^{(0)}_{\Sigma^+}(u_{\dw})={1\over 3}~,~~~
n^{(0)}_{\Sigma^+}(s_{\up})={1\over 3}~,~~~
n^{(0)}_{\Sigma^+}(s_{\dw})={2\over 3}~.
\eqno (16)$$
Using (5) (with replacement, $p\to \Sigma^+$ and $\sum\limits_{q=u,d} 
\to \sum\limits_{q=u,s}$), (16), and Table I, we can obtain
$\Delta u^{\Sigma^+}$, $\Delta d^{\Sigma^+}$, and $\Delta s^{\Sigma^+}$.
Similarly, we can obtain results for $\Sigma^0$, $\Lambda^0$, and
$\Xi^0$. Those for $\Sigma^-$, and $\Xi^-$, can be obtained by using 
the isospin symmetry relations. The analytical expressions of all $\Delta
q^{B}$ are listed in the upper half of Table VII. A few remarks are in 
order.
\begin{itemize}
\item{The total spin content of quarks and antiquarks in the octet 
baryons can be written as 
$$\sum\limits_{q}<s_z>_{q+\bar q}^{B}={1\over 2}-{a\over 3}
(c_1\xi_1+c_2\xi_2),
\eqno (17)$$
where $\xi_2\equiv 2\epsilon+f_s$ [definition of $f_s$ see Eq.(6b)] and 
$(c_1, c_2)$=(3, 0), $(4, -1)$, (0, 3), and $(-1, 4)$ for B=N, $\Sigma$, 
$\Lambda$, and $\Xi$ respectively. One can see that the spin reductions 
for all members in the {\it same} isospin multiplet are the {\it same}, 
but they are different for different isospin multiplets, except for the
SU(3)-symmetry limit ($\xi_1=\xi_2=2+f$) and U(3)-symmetry limit 
($\xi_1=\xi_2=3$, in this case, 
$\sum\limits_{q}<s_z>_{q+\bar q}^{N,\Sigma,\Lambda,\Xi}=
{1\over 2}-3a$). }
\item{Using parameters $\xi_1\simeq 2.07$, and $\xi_2\simeq
1.27$, we plot the quark and antiquark spin content in different octet
baryons as functions of the parameter $a$ in Fig.3. Taking $a\simeq
0.145$, one obtains
$$\sum\limits_{q}<s_z>_{q+\bar q}^{B}\simeq 0.20,~0.16,~0.32,~0.35
~~~~(B=N,~\Sigma,~\Lambda,~\Xi).
\eqno (18)$$
}
\item{The spin of the $\Lambda$: In the naive SU(3) symmetric quark 
model, all the spin of the $\Lambda$ comes from the spin of the $s$ 
quark
$$\Delta u^{\Lambda}=\Delta d^{\Lambda}=0,~~~\Delta s^{\Lambda}=1,~~~
\sum\limits_q\Delta q^{\Lambda}=1.
$$
From Table VIII, the chiral quark model predicts
$$\Delta u^{\Lambda}=\Delta d^{\Lambda}=-0.07,~~~\Delta s^{\Lambda}=0.77,
~~~\sum\limits_q\Delta q^{\Lambda}=0.64.
$$
This can compare with the result, $\Delta u^{\Lambda}=\Delta d^{\Lambda}
=-0.23$, $\Delta s^{\Lambda}=0.58$, and $\sum\limits_q\Delta 
q^{\Lambda}=0.12$ given in \cite{jaffe96}. Our result
indicates that about $64\%$ of the $\Lambda$ spin is contributed
by the spin of quarks, which include $s$ quark (positive, +$77\%$) and 
$u,d$ quarks (negative, total $-13\%$). The remaining part ($36\%$) 
comes from the orbital angular momentum of quarks and antiquarks in 
our model and possibly from the gluon contributions in a modified
chiral quark model \cite{song9804461} with the gluon mixing \cite{lip}. }
\end{itemize}

\leftline{\quad (b)~\bf Orbital angular momentum in octet baryons}

Similar to the nucleon case, the orbital angular momentum carried 
by quarks and antiquarks in other octet baryons can be calculated.
The analytical results for different isospin multiplets are listed 
in the lower half of Table VII. 
The total orbital angular momentum carried by all quarks and 
and antiquarks in the baryon $B$ can be written as
$$\sum\limits_{q}<L_z>_{q+\bar q}^{B}={a\over 3}(c_1\xi_1+c_2\xi_2),
\eqno (19)$$
where $c_1$ and $c_2$ are the same as defined in Eq. (17). The sum of 
spin (17) and orbital angular momentum (19) gives
$$\sum\limits_{q}[<s_z>_{q+\bar q}^{B}+
<L_z>_{q+\bar q}^{B}]={1\over 2},~~~~(B=N,~\Sigma,~\Lambda,~
\Xi)
\eqno (20)$$
Once again, the loss of the quark spin is {\it compensated} by the gain 
of the orbital motion of quarks and antiquarks. Results and conclusions
obtained in section III for the nucleon hold for other octet baryons 
as well. Numerical results of the spin and orbital angular momentum 
for each quark flavor in the baryon octet are listed in Table VIII. 
We make some remarks here.
\begin{itemize}
\item{From (18) and (19), one can see that nearly $68\%$ ($36\%$, 
$30\%$) of $\Sigma^+$ ($\Lambda$, $\Xi^0$) spin is coming from the 
orbital motion of quarks and antiquarks, and $32\%$ ($64\%$, $70\%$) 
is contributed by the quark and antiquark spins. This result does not
depend on the partition factor $\kappa$.}
\item{Using relations $<L_z>_{u,d}^{\Sigma^-,\Xi^-}=<L_z>_{d,u}^{\Sigma^+,
\Xi^0}$ and $<L_z>_s^{\Sigma^-,\Xi^-}=<L_z>_s^{\Sigma^+,\Xi^0}$ from
the isospin symmetry, one can obtain orbital angular momenta in
$\Sigma^-$ and $\Xi^-$. Similar to the nucleon case, we have 
$<L_z>^{\Sigma^+,\Xi^0}_d=<L_z>^{\Sigma^+,\Xi^0}_{\bar d}$ at
$\kappa=1/3$. This is because there is no valence $d$ quark in 
$\Sigma^+$ and $\Xi^0$, hence all $d$ and $\bar d$ must come from the GB.}
\item{Similar to the nucleon case [see $\Delta_3$
expression below Eq.(7c)], we have a general formula for the octet baryons
$$\Delta u^B-\Delta d^B=c_B[1-(\epsilon+2f)],
\eqno (21)$$
where $c_B=5/3$, $4/3$, and $-1/3$ for $B=p$, $\Sigma^+$, and
$\Xi^0$ respectively.}
\item{It is easy to check from the analytic results given in Table VII 
that the following identity holds for the $u-$quark spin
$$\Delta u^p-\Delta u^n+\Delta u^{\Sigma^-}-\Delta u^{\Sigma^+}+
\Delta u^{\Xi^0}-\Delta u^{\Xi^-}=0,
\eqno (22a)$$
and the same identity holds for the $d-$quark and $s-$quark spins. 
One can also show by explicit calculation that the orbital angular 
momentums $<L_z>_u^B$ in the octet baryons satisfy similar
identity
$$<L_z>_u^p-<L_z>_u^n+<L_z>_u^{\Sigma^-}-<L_z>_u^{\Sigma^+}+
<L_z>_u^{\Xi^0}-<L_z>_u^{\Xi^-}=0
\eqno (22b)$$
Similar relations hold for $<L_z>_{d}^B$, $<L_z>_{s}^B$, and
$<L_z>_{\bar q}^B$. Combining (22a) and (22b), one obtains a sum rule
for the magnetic moments [see Eq.(25) below] in the baryon octet 
$$\mu_p-\mu_n+\mu_{\Sigma^-}-\mu_{\Sigma^+}+
\mu_{\Xi^0}-\mu_{\Xi^-}=0.
\eqno (22c)$$
This sum rule was discussed in \cite{los} without the orbital 
contributions. Our result shows that the sum rule (22c) holds 
in the symmetry breaking chiral quark model even the orbital 
contributions are included. Quark spin content, without the
orbital angular momentum, in octet baryons were discussed in 
\cite{los,wb}.}
\end{itemize}
\smallskip

\leftline{\quad (c)~\bf Decuplet baryons}

The above discussion can be extended to the baryon decuplet. The
analytical results are listed in Table IX. The explicit calculation 
shows that
$$(\Delta u)^{\Delta^{-}}=(\Delta d)^{\Delta^{++}},~ 
(\Delta u)^{\Delta^{0}}=(\Delta d)^{\Delta^{+}},~
(\Delta u)^{\Sigma^{*-}}=(\Delta d)^{\Sigma^{*+}},~ 
(\Delta u)^{\Xi^{*-}}=(\Delta d)^{\Xi^{*0}}.
$$
They are due to the isospin symmetry of the decuplet baryon wave 
functions. Therefore, in Table IX, we only list results for $\Delta^{++}$, 
$\Delta^{+}$, $\Sigma^{*+}$, $\Sigma^{*0}$, $\Xi^{*0}$, and $\Omega^{-}$. 
A few remarks are in order.
\begin{itemize}
\item{From Table IX, one obtains
$$\sum\limits_{q}<s_z>_{q+\bar q}^{B^*}={3\over
2}-a[3\xi_1+S(\xi_1-\xi_2)],
\eqno (23a)$$
where the $S$ is the {\it strangeness} quantum number of the decuplet 
baryon $B^*$. Eq.(23a) leads to an {\it equal spacing rule} for the total 
quark spin in the baryon decuplet, and the {\it spacing} is 
$$\sum\limits_{q}<s_z>_{q+\bar q}^{\Omega-\Xi}=
\sum\limits_{q}<s_z>_{q+\bar q}^{\Xi-\Sigma}=
\sum\limits_{q}<s_z>_{q+\bar q}^{\Sigma-\Delta}=
a(\xi_1-\xi_2).
\eqno (23b)$$
The equal spacing rule (23b) was also discussed in \cite{los} without 
the orbital contributions.}
\item{From (23a), for the strangeless $\Delta$ multiplet, $S=0$, one
obtains
$$\sum\limits_{q}<s_z>_{q+\bar q}^{\Delta}=3[{1\over 2}-{a\over
3}(3\xi_1)]=3\sum\limits_{q}<s_z>_{q+\bar q}^N,
\eqno (23c)$$
i.e. total spin content of $\Delta$ baryon is {\it three} times that of 
the nucleon, this is a reasonable result.} 
\item{From Table IX, the total orbital angular momentum carried by quarks
and antiquarks can be written as
$$\sum\limits_{q}<L_z>_{q+\bar q}^{B^*}=a[3\xi_1+S(\xi_1-\xi_2)].
\eqno (24a)$$
Similar to the total quark spin case, Eq.(24a) leads to an {\it equal 
spacing rule} for the orbital angular momentum. The spacing is
$$\sum\limits_{q}<L_z>_{q+\bar q}^{\Omega-\Xi}=
\sum\limits_{q}<L_z>_{q+\bar q}^{\Xi-\Sigma}=
\sum\limits_{q}<L_z>_{q+\bar q}^{\Sigma-\Delta}=
-a(\xi_1-\xi_2), 
\eqno (24b)$$
which is the same as (23b) but with {\it opposite} sign. 
The sum of spin (23a) and orbital angular momentum (24a) gives
$$\sum\limits_{q}[<L_z>_{q+\bar q}^{B^*}+<s_z>_{q+\bar q}^{B^*}]
={3\over 2}.
\eqno (24c)$$
Once again, the spin reduction is compensated by the increase of
orbital angular momentum and keep the total angular momentum of the 
baryon ({\it now is {3/2} for the decuplet} !) unchanged.}
\end{itemize} 
\bigskip

\leftline{\bf V. Baryon Magnetic Moments}

The baryon magnetic moment depends on both spin and orbital motions 
of quarks and antiquarks. In the chiral quark model all 
antiquark sea polarizations are zero, the baryon magnetic moment
can be written as
$$\mu_{B(B^*)}=\sum\limits_{q}\mu_q[(\Delta q)^{B(B^*)}
+<L_z>^{B(B^*)}_{q}-<L_z>^{B(B^*)}_{\bar q}]
\equiv\sum\limits_{q}\mu_qC_q^B ,
\eqno (25)$$
where $B(B^*)$ denote the octet (decuplet) baryons and $\mu_q$s are 
the magnetic moments of quarks. We have assumed that the magnetic 
moment of the baryon is the sum of spin and orbital magnetic moments 
of individual charged particles (quarks or antiquarks). The assumption 
of the {\it additivity} is commonly believed to be a good approximation 
for a loosely bound system, which is the basic description for the 
baryon in the chiral quark model. In addition, the baryon may contain 
other neutral particles, such as gluons (e.g. see discussion in 
\cite{bs90}). Although the gluon does not make any contribution to 
the magnetic moment, the existence of intrinsic gluon would change 
the valence quark structure of the baryon due to the spin and color 
couplings between the gluon and quarks. In this paper, we assume that 
the gluon effect is small and will be neglected in our calculation.
The quark spin and orbital angular momentum in a modified chiral quark 
model with the gluon mixing has been discuused in \cite{song9804461}.
 
To calculate the baryon magnetic moment, we need to know the spin 
content $\Delta q$ (note that $\Delta\bar q=0$ in the chiral quark 
model) and the {\it difference} between the orbital angular momentum 
carried by the quark $q$ and that carried by corresponding antiquark 
$\bar q$. This difference is denoted by $<L_z>^{B}_{q-\bar q}\equiv 
<L_z>^{B}_{q}-<L_z>^{B}_{\bar q}$. For example, one has for the 
$u$-quark 
$$<L_z>_{u-\bar u}^B=\sum\limits_{q}
[n^{(0)}_B(q_{\up})-n^{(0)}_B(q_{\dw})][<L_z>_{u/q_{\up}}-
<L_z>_{{\bar u}/q_{\up}}].
\eqno (26)$$
Similar equations hold for the $d$-quark, $s$-quark, and corresponding 
antiquarks.
\smallskip

\leftline{\quad (a)~\bf Octet baryons}

Substituting (7a) and (26) into (25), one can obtain the analytic 
expressions of the magnetic moments for the octet baryons. It is easy 
to verify that they satisfy the following sum rules 
$$(4.70)~~~{\mu}_p-{\mu}_n={\mu}_{\Sigma^+}-{\mu}_{\Sigma^-}-
({\mu}_{\Xi^0}-{\mu}_{\Xi^-})~~~(4.22),
\eqno (27a)$$
$$(3.66)~~~-6{\mu}_{\Lambda}=-2({\mu}_p+{\mu}_n+{\mu}_{\Xi^0}+{\mu}_{\Xi^-})
+({\mu}_{\Sigma^+}+{\mu}_{\Sigma^-})~~~(3.34),
\eqno (27b)$$
$$(4.14)~~~{\mu}_p^2-{\mu}_n^2=({\mu}_{\Sigma^+}^2-{\mu}_{\Sigma^-}^2)
-({\mu}_{\Xi^0}^2-{\mu}_{\Xi^-}^2)~~~(3.56),
\eqno (27c)$$
$$(0.33)~~~{\mu}_p-{\mu}_{\Sigma^+}={3\over
5}({\mu}_{\Sigma^-}-{\mu}_{\Xi^-})
-({\mu}_n-{\mu}_{\Xi^0})~~~(0.31),
\eqno (27d)$$
where the numerical values shown in parentheses are taken from the data
\cite{pdg00}. Relations (27a) [or (22c)] and (27b) were
first given by Franklin in \cite{frank}. Linear sum rules (27a), (27b), 
and nonlinear sum rule (27c) are not new and violated at about $10-15\%$ 
level. They have been discussed in many works, for instance
\cite{br,karl,sg}. However, the new relation (27d) is rather well 
satisfied. Our result shows that if the $SU(3)\otimes SU(2)$ valence 
quark structure is used, the chiral fluctuation cannot change these sum 
rules even orbital contributions are included. We note that the sum rules 
(27a-27c) also hold for more general case \cite{sg}.

Numerical results of applying Eq.(25) to the magnetic moments of the
baryon octet are listed in Table X. We find that if we choose parameters 
$\mu_u$, $\mu_d$ and $\mu_s$ by fitting the measured values of
$\mu_p$, $\mu_n$ and $\mu_\Lambda$ as is also done in the simple SU(6) 
quark model (NQM) \cite{pdg00}, then results for all different $\kappa$
values are completely $identical$ with the NQM results. Three remarks
are in order.
\begin{itemize} 
\item{According to the fitting procedure, for {\it a given set} of 
$\mu_{u,d,s}^{(0)}$ in the NQM, there is {\it a
corresponding set} of $\mu_{u,d,s}$ which gives the same values of
$\mu_p$, $\mu_n$ and $\mu_\Lambda$ in the chiral quark model.}

\item{Seven magnetic moments of the baryon octet satisfy four sum rules,
Eqs.(27a-27d), which hold for both the chiral quark model and the NQM
and appear to be reasonably well satisfied. Hence, as soon as three 
baryon magnetic moments are taken to be the same in both cases, the 
remaining four should agree as well. Hence, effects of the chiral 
splitting with or without the orbital contribution can be absorbed 
into the quark magnetic moments. It implies that for describing the 
baryon magnetic moments, the NQM is {\it as good as} the chiral quark 
model. This is similar to the conclusion given in \cite{cl3}. Our 
result shows that the magnetic moment is presumably {\it not} a 
suitable observable for revealing the quark orbital angular momentum 
in the nucleon, {\it unless the quark magnetic moments are 
given.}}

\item{We have shown in section III, the total amount of quark spin 
reduction is canceled by the equal amount increase of the quark orbital 
angular momentum $<L_z>_{q+\bar q}$, but this exact cancellation does 
not occur for each quark flavor. To see whether similar cancellation 
occurs in the baryon magnetic moments, we turn into Eq.(25). The 
relevant quantity is $C_q^B\equiv\Delta q^B+<L_z>_{q-\bar q}^B$ defined 
in Eq.(25). Using superscripts (0), (1), and (2) to denote the quantity
in NQM, $\chi$QM with, and without $<L_z>_{q-\bar q}$ respectively,
we have 
$$C_u^{(0)p}=4/3,\quad C_d^{(0)p}=-1/3,\quad C_s^{(0)p}=0,~~~~
({\rm NQM},~<L_z>_{q}=<L_z>_{\bar q}=0),
\eqno (28a)$$
$$C_u^{(1)p}=0.996,~C_d^{(1)p}=-0.403,~ C_s^{(1)p}=-0.067,~
(\chi{\rm QM},~{\rm with}~~<L_z>_{q-\bar q}),
\eqno (28b)$$
$$C_u^{(2)p}=0.863,~C_d^{(2)p}=-0.397,~ C_s^{(2)p}=-0.067,~ 
(\chi{\rm QM},~{\rm without}~<L_z>_{q-\bar q}).
\eqno (28c)
$$
The orbital contribution does move $C_u^{(2)p}$ up to $C_u^{(1)p}$, 
but it is still far below 4/3. For the $d$-quark flavor, the orbital
contribution moves $C_d^{(2)p}$ down and $C_d^{(1)p}$ is even more 
negative than $-1/3$. Comparing (28a-28c) with (15a-15c), one can see 
that in the proton magnetic moment, there is a {\it partial cancellation}
between the contributions from the quark spin reduction and the quark 
orbital angular momentum for the $u$-flavor ($<L_z>^p_{u-\bar u}=0.133$), 
but {\it no cancellation} for the $s$-flavor ($<L_z>^p_{s-\bar s}=0$). 
For the $d$-flavor, the orbital contribution even {\it enhances} the
effect of the spin reduction from the chiral splitting ($<L_z>_{d-\bar
d}=-0.027$). Hence the so called `cancellation' here is quite {\it
different} from the total angular momentum case. This difference is 
due to the orbital contribution in the magnetic moment (25) is
$<L_z>_{q-\bar q}$, but in the nucleon spin sum rule (10d) that is 
$<L_z>_{q+\bar q}$.}
\end{itemize}
\smallskip
 
\leftline{\quad (b)~\bf Decuplet baryons}
 
Similar to the octet baryons, the decuplet magnetic moments are 
calculated and the numerical results are listed in Table XI. It
should be noted that we use the $same$ set of $\mu_u$, $\mu_d$ 
and $\mu_s$ as in the octet sector, i.e. {\it no} new parameters
are introduced. 

It is easy to verify from the analytic results given in Table IX that 
the following {\it equal spacing} rules hold for the decuplet magnetic 
moments,
$$\mu_{\Delta^{++}}-\mu_{\Delta^{+}}=
\mu_{\Delta^{+}}-\mu_{\Delta^{0}}=
\mu_{\Delta^{0}}-\mu_{\Delta^{-}}=
\mu_{\Sigma^{*+}}-\mu_{\Sigma^{*0}}=
\mu_{\Sigma^{*0}}-\mu_{\Sigma^{*-}}=
\mu_{\Xi^{*0}}-\mu_{\Xi^{*-}},
\eqno (29a)$$
$$\mu_{\Delta^{+}}-\mu_{\Sigma^{*+}}=
\mu_{\Delta^{0}}-\mu_{\Sigma^{*0}}=
\mu_{\Delta^{-}}-\mu_{\Sigma^{*-}}=
\mu_{\Sigma^{*0}}-\mu_{\Xi^{*0}}=
\mu_{\Sigma^{*-}}-\mu_{\Xi^{*-}}=
\mu_{\Xi^{*-}}-\mu_{\Omega^{*-}}.
\eqno (29b)$$
The spacing in Eqs.(29a) and (29b) are 2.82 n.m. and $-0.36$ n.m.
respectively. Our results also show that the decuplet magnetic 
moments with or without orbital contributions are {\it approximately 
the same} provided the quark magnetic moments change accordingly. 
\bigskip

\leftline{\bf VI. Discussions and Summary}
\smallskip

For the quark flavor and spin observables of the nucleon, the predictions 
of the chiral quark model are in good agreement with the existing data
(see Table IV). We now consider different scenarios of the spin sum
rule with or without the gluon contribution. 
\begin{itemize}
\item{Assuming $<J_z>_G\simeq 0$, our result shows
$\sum\limits_{q}<L_z>^p_{q+\bar q}\simeq 0.30$, i.e. $60\%$ of 
the proton spin is contributed by the orbital motion of quarks and 
antiquarks. The nucleon spin sum rule (1) becomes 
$$ {1\over 2}~=~\sum\limits_q<s_z>_{q+\bar q}~(\sim 0.2)~+~
\sum\limits_q<L_z>^p_{q+\bar q}~(\sim 0.3)~+~
<J_z>_{G}~(\sim 0).
\eqno(30a)$$}
\item{However, if $<J_z>_G$ is {\it nonzero} at low $Q^2$ scale as
suggested in \cite{bj}, results given in sections III and IV should 
be modified. Using the nucleon wave function given in the gluon mixing 
model \cite{lip} and considering the chiral splitting mechanism, we
have shown \cite{song9804461} that the ratio of 
$\sum\limits_q<s_z>^p_{q+\bar q}$ and $\sum\limits_q<L_z>^p_{q+\bar q}$ 
is {\it the same as} that given in (11b). For the parameter set given 
in (12), this ratio $1/(2\xi_1a)-1$ is approximately equal to 2/3.
Assuming $<J_z>_G(1~{GeV^2})\simeq 0.15$, we have 
$${1\over 2}~=~\sum\limits_{q}<s_z>_{q+\bar q}~(\sim 0.14)~+~
\sum\limits_{q}<L_z>_{q+\bar q}~(\sim 0.21)~+~
<J_z>_G~(\sim 0.15).
\eqno (30b)$$}
\item{The quantities calculated in the chiral quark model should be 
viewed as the physical observables at the $Q^2$ range lower than 1
GeV$^2$. On the other hand, at high $Q^2$ ($\geq 1-2$ GeV$^2$), the QCD
parton model works and the main degrees of freedom are quarks and gluons.
According to \cite{jth96}, in the large $Q^2$ limit, the partition of
the nucleon spin between quarks and gluons follows the well-known
partition of the nucleon momentum. It implies that   
$\sum\limits_{q}<J_z>_{q+\bar q}\simeq <J_z>_G\simeq 0.25$, i.e.
the nucleon spin is approximately {\it equally} shared by quarks 
(antiquarks) and gluons in the large $Q^2$ limit. An interesting question
is whether this partition rule holds at the low $Q^2$ scale ? We do not
know the answer theoretically because it is difficult to solve QCD at this
scale. Assuming the partition rule holds at the low $Q^2$, then we have
third scenario
$${1\over 2}~=~\sum\limits_{q}<s_z>_{q+\bar q}~(\sim 0.10)~+~
\sum\limits_{q}<L_z>_{q+\bar q}~(\sim 0.15)~+~
<J_z>_G~(\sim 0.25),
\eqno (30c)$$ 
Perhaps experiments in the transition area $Q^2\simeq 1-2$ GeV$^2$
will tell us which scenario is closer to the reality.}  
\end{itemize}

In this paper, we discussed the quark flavor, spin and orbital contents 
in the baryon in the chiral quark model. Contrary to the reduction effect
on the quark spin component, the chiral splitting mechanism produces a 
positive orbital angular momentum shared by quarks and antiquarks. 
Analytical and numerical results for the spin and orbital contents 
carried by different quark flavors in the baryon are presented. 
Attention has been drawn to the cancellation between the spin and 
orbital contributions in the nucleon spin sum rule and in the baryon 
magnetic moments. 

In summary, the chiral quark model with only $a$ $few$ parameters can well
explain many nucleon properties: (1) strong flavor asymmetry of light 
antiquark sea: $\bar d> \bar u$, (2) nonzero strange quark content, 
$<\bar ss>\neq 0$, (3) sum of quark spins is small, $\sum\limits_q
<s_z>_{q+\bar q}\simeq 0.1-0.2$, (4) sea antiquarks are not polarized: 
$\Delta\bar q\simeq 0$ ($q=u,d,...$), (5) polarizations of the sea quarks 
are nonzero and negative, $\Delta q_{sea}< 0$, and (6) the orbital angular 
momentum of the sea quark is parallel to the proton spin. (1)-(4) are 
consistent with data (see Table IV), and (5)-(6) could be tested by 
experiments in the near future.

\bigskip

\leftline{\bf Acknowledgments}

I would like to thank Xiangdong Ji, and P. K. Kabir for useful comments 
and suggestions. This work was supported in part by the U. S. Department 
of Energy and the Institute of Nuclear and Particle Physics, Department 
of Physics, University of Virginia.
\bigskip


\bigskip

\begin{table}[ht]
\begin{center}
\caption{The probabilities $P_{q_{\up}}(q'_{\up,\dw},\bar q'_{\up,\dw})$
and $P_{q_{\up}}(q'_{\up,\dw},{\bar q}'_{\up,\dw})$} 
\begin{tabular}{cccc} 
\hline
$q'$ &$P_{u_{\up}}(q'_{\up,\dw})$ & $P_{d_{\up}}(q'_{\up,\dw})$ &
$P_{s_{\up}}(q'_{\up,\dw})$ \\ 
\hline 
$u_{\up}$ & $1-({{1+\epsilon}\over 2}+f)a+
{a\over {18}}(3-A)^2$ & ${a\over {18}}A^2$ &${a\over {18}}B^2$ \\
$u_{\dw}$ & $({{1+\epsilon}\over 2}+f)a+
{a\over {18}}(3-A)^2$ & $a+{a\over {18}}A^2$ &$\epsilon a+{a\over
{18}}B^2$ \\
$d_{\up}$ & ${a\over {18}}A^2$ &$1-({{1+\epsilon}\over 2}+f)a+
 {a\over {18}}(3-A)^2$ &${a\over {18}}B^2$ \\
$d_{\dw}$ & $a+{a\over {18}}A^2$ &
$({{1+\epsilon}\over 2}+f)a+{a\over {18}}(3-A)^2$ & 
$\epsilon a+{a\over {18}}B^2$ \\
$s_{\up}$ & ${a\over {18}}B^2$ &${a\over {18}}B^2$ &
$1-(\epsilon+f_s)a+{a\over {18}}C^2$ \\
$s_{\dw}$ & $\epsilon a+{a\over {18}}B^2$ & $\epsilon a+{a\over {18}}B^2$ 
&$(\epsilon+f_s)a+{a\over {18}}C^2$ \\
\hline
${\bar u}_{\up,\dw}$ &${a\over {18}}(3-A)^2$ & ${a\over 2}+{a\over
{18}}A^2$ &${{\epsilon a}\over 2}+{a\over {18}}B^2$ \\
${\bar d}_{\up,\dw}$ &${a\over 2}+{a\over {18}}A^2$&
${a\over {18}}(3-A)^2$ &${{\epsilon a}\over 2}+{a\over {18}}B^2$ \\
${\bar s}_{\up,\dw}$ &${{\epsilon a}\over 2}+{a\over {18}}B^2$&
${{\epsilon a}\over 2}+{a\over {18}}B^2$&
${a\over {18}}C^2$ \\
\hline
\end{tabular}
\end{center}
\end{table}

\begin{table}[ht]
\begin{center}
\caption{The spin-up, spin-down quark (antiquark), 
spin-average and spin-weighted quark (antiquark) content in the proton.
Where $U_1\equiv{1\over 3}[A^2+2(3-A)^2]$, $D_1\equiv{1\over
3}[2A^2+(3-A)^2]$, and $U_2=4D_2\equiv 4(\epsilon+2f-1)$.}
\begin{tabular}{ccc} 
\hline
$u_{\up}={5\over 3}+{a\over 3}(-2+{{U_1}\over 2}-{{U_2}\over 2})$& 
$d_{\up}={1\over 3}+{a\over 3}(2+{{D_1}\over 2}+{{D_2}\over 2})$& 
$s_{\up}=\epsilon a+{a\over 3}({{B^2}\over 2})$\\
$u_{\dw}={1\over 3}+{a\over 3}(5+{{U_1}\over 2}+{{U_2}\over 2})$& 
$d_{\dw}={2\over 3}+{a\over 3}(4+{{D_1}\over 2}-{{D_2}\over 2})$& 
$s_{\dw}=2\epsilon a+{a\over 3}({{B^2}\over 2})$\\
\hline 
${\bar u}_{\up}={\bar u}_{\dw}={a\over 2}+{a\over 3}({{U_1}\over 2})$&
${\bar d}_{\up}={\bar d}_{\dw}=a+{a\over 3}({{D_1}\over 2})$ &
${\bar s}_{\up}={\bar s}_{\dw}={{3\epsilon a}\over 2}+{a\over
3}({{B^2}\over 2})$\\
\hline
\hline
$u=2+{a\over 3}(3+U_1)$&$d=1+{a\over 3}(6+D_1)$& $s=3\epsilon a+{a\over
3}B^2$\\
\hline
$\bar u={a\over 3}(3+U_1)$&$\bar d={a\over 3}(6+D_1)$& $\bar s=3\epsilon
a+{a\over 3}B^2$\\
\hline
\hline
$\Delta u={4\over 3}[1-a(\epsilon+2f)]-a$ &$\Delta d={{-1}\over
3}[1-a(\epsilon+2f)]-a$ & $\Delta s=a(1-\epsilon)-a$ \\
\hline
$\Delta{\bar u}=0$ & $\Delta{\bar d}=0$ &$\Delta{\bar s}=0$ \\
\hline
\end{tabular}
\end{center}
\end{table}

\begin{table}[ht]
\begin{center}
\caption{The orbital angular momentum carried by the quark $q'$ 
($\bar q'$), {\it spin-up and -down are included}, 
from a valence spin-up quark $q_{\up}$ fluctuates into all 
allowed final states. Where $\delta\equiv (1-3\kappa)/\kappa$.}
\begin{tabular}{cccc} 
&$<L_z>_{q',{\bar q'}/u_{\up}}$ &$<L_z>_{q',{\bar q'}/d_{\up}}$ 
&$<L_z>_{q',{\bar q'}/s_{\up}}$\\ 
\hline 
$q'=u$ & $\kappa a[\xi_1+f\delta+{{(3-A)^2}\over 9}]$ &
$\kappa a[1+\delta+{{A^2}\over 9}]$ 
&$\kappa a[\epsilon(1+\delta)+{{B^2}\over {9}}]$ \\
$q'=d$ & $\kappa a[1+\delta+{{A^2}\over {9}}]$
&$\kappa a[\xi_1+f\delta+{{(3-A)^2}\over {9}}]$
&$\kappa a[\epsilon(1+\delta)+{{B^2}\over {9}}]$ \\
$q'=s$ &$\kappa a[\epsilon(1+\delta)+{{B^2}\over {9}}]$ 
&$\kappa a[\epsilon(1+\delta)+{{B^2}\over {9}}]$ &
$\kappa a[\xi_2+f_s\delta+{{C^2}\over {9}}]$\\
\hline
${\bar q'}={\bar u}$ &$\kappa a[{{(3-A)^2}\over {9}}]$
&$\kappa a[1+{{A^2}\over {9}}]$ 
&$\kappa a[\epsilon+{{B^2}\over {9}}]$ \\
${\bar q'}={\bar d}$ 
&$\kappa a[1+{{A^2}\over {9}}]$&$\kappa a[{{(3-A)^2}\over {9}}]$
&$\kappa a[\epsilon+{{B^2}\over {9}}]$ \\
${\bar q'}={\bar s}$&$\kappa a[\epsilon+{{B^2}\over {9}}]$ 
&$\kappa a[\epsilon+{{B^2}\over {9}}]$ 
&$\kappa a[{{C^2}\over {9}}]$ \\
\hline
\end{tabular}
\end{center}
\end{table}

\begin{table}[ht]
\begin{center}
\caption{Quark spin and flavor observables in the proton. The quantity
used as input is indicated by a star. }
\begin{tabular}{cccc} 
\hline
Quantity & Data& This paper &NQM\\
\hline 
$\bar d-\bar u$ & $0.147\pm 0.039$ \cite{nmc} & $0.143^*$  &0\\
& $0.100\pm 0.018$ \cite{e866} & &  \\
${{\bar u}/{\bar d}}$ & $[{{\bar u(x)}\over {\bar d(x)}}]_{x=0.18}=0.51\pm
0.06$ \cite{na51}& 0.64 &   $-$\\
 & $[{{\bar u(x)}\over {\bar d(x)}}]_{0.1<x<0.2}=0.67\pm 0.06$ \cite{e866}
& &\\
${{2\bar s}/{(\bar u+\bar d)}}$ & ${{<2x\bar s(x)>}\over {<x(\bar
u(x)+\bar d(x))>}}=0.477\pm 0.051$ \cite{ccfr95}& 0.72 & $-$\\
 ${{2\bar s}/{(u+d)}}$ & ${{<2x\bar s(x)>}\over
{<x(u(x)+d(x))>}}=0.099\pm0.009$ \cite{ccfr95} & 0.13 &0\\
 ${{\sum\bar q}/{\sum q}}$ & ${{\sum<x\bar q(x)>}\over{\sum<xq(x)>}}
=0.245\pm 0.005$ \cite{ccfr95} & 0.23 &0\\
 $f_s$ & $0.10\pm 0.06$ \cite{gls91} & 0.10 & 0\\
       & $0.15\pm 0.03$ \cite{dll95} &      &    \\
       & ${{<2x\bar s(x)>}\over {\sum<x(q(x)+\bar q(x))>}}
=0.076\pm 0.022$ \cite{ccfr95} & &  \\
$f_3/f_8$ & $0.21\pm 0.05$ \cite{cl1} & 0.22 & 1/3\\
\hline 
$\Delta u$ & $0.85\pm 0.05$ \cite{smc97} & 0.86 & 4/3\\
$\Delta d$&$-0.41\pm 0.05$ \cite{smc97} &$-$0.40&$-$1/3\\
$\Delta s$&$-0.07\pm 0.05$ \cite{smc97} &$-0.07$&0\\
$\Delta\bar u$, $\Delta\bar d$ & $-0.02\pm 0.11$ \cite{smc96} &0&0 \\
$\Delta_3/\Delta_8$ &2.17$\pm 0.10$&2.12& 5/3\\
$\Delta_3$ &1.2601$\pm 0.0028$ \cite{pdg00}&1.26$^{*}$& 5/3\\
$\Delta_8$& 0.579$\pm 0.025$ \cite{pdg00}& 0.60$^{*}$&1 \\
\hline
\end{tabular}
\end{center}
\end{table}

\begin{table}[ht]
\begin{center}
\caption{Quark spin and orbital angular momentum in the proton in
different models.}
\begin{tabular}{ccccccc} 
Quantity & Data \cite{smc97}   && This paper& &  Sehgal \cite{sehgal} & NQM\\ 
&& $\kappa=1/4$ & $\kappa=1/3$& $\kappa=3/8$ &&\\
\hline 
$<L_z>_u^p$       & $-$ & 0.115   & 0.130   & 0.138  & $-$    & 0  \\ 
$<L_z>_d^p$       & $-$ & 0.073   & 0.043   & 0.027  & $-$    & 0  \\ 
$<L_z>_s^p$       & $-$ & 0.038   & 0.028   & 0.023  & $-$    & 0  \\ 
$<L_z>_{\bar u}^p$& $-$ &$-$0.003 & $-$0.003&$-$0.004 & $-$    & 0  \\ 
$<L_z>_{\bar d}^p$& $-$ & 0.057   & 0.076   & 0.086  & $-$    & 0  \\ 
$<L_z>_{\bar s}^p$& $-$ & 0.021   & 0.028   & 0.031  & $-$    & 0  \\ 
\hline
$\sum\limits_q<L_z>_{q+\bar q}^p$ & $-$ & 0.30 & 0.30& 0.30 &0.39& 0  \\ 
\hline
$\Delta u^p$ & $0.85\pm 0.05$ & & 0.86     &  & 0.78    & 4/3  \\ 
$\Delta d^p$ & $-0.41\pm 0.05$ & & $-0.40$ &  & $-0.34$ & $-1/3$\\ 
$\Delta s^p$ & $-0.07\pm 0.05$ & &$-0.07$ &  & $-0.14$ & 0  \\ 
\hline
${1\over 2}\Delta\Sigma^p$ & $0.19\pm 0.06$ & & 0.20 &  & 0.08 & 1/2\\ 
\hline
\end{tabular}
\end{center}
\end{table}

\begin{table}[ht]
\begin{center}
\caption{Quark spin and orbital angular momentum in different models.}
\begin{tabular}{cccccc} 
\hline 
 & NQM & MIT bag & This paper & CS \cite{cs97}&Skyrme\\ 
\hline 
$\sum\limits_q<s_z>^p_{q+\bar q}$ &1/2& 0.32 & 0.20 & 0.08&0\\
\hline 
$\sum\limits_q<L_z>^p_{q+\bar q}$ & 0 & 0.18 & 0.30 & 0.42& 1/2\\
\hline
\end{tabular}
\end{center}
\end{table}

\begin{table}[ht]
\begin{center}
\caption{The quark spin and orbital content in the octet baryons. 
Where $\xi_1=1+\epsilon+f$, $\xi_2=2\epsilon+f_s$, and 
$\delta\equiv (1-3\kappa)/\kappa$.}
\begin{tabular}{cccc} 
 Baryon &$\Delta u^B$ & $\Delta d^B$& $\Delta s^B$ \\
\hline 
p & ${4\over 3}-{a\over 3}(8\xi_1-4\epsilon-5)$& 
$-{1\over 3}-{a\over 3}(-2\xi_1+\epsilon+5)$ & $-a\epsilon$\\
$\Sigma^+$ & ${4\over 3}-{a\over 3}(8\xi_1-5\epsilon-4)$& 
$-{a\over 3}(4-\epsilon)$ & $-{1\over 3}-{{2a}\over 3}(3\epsilon-\xi_2)$\\
$\Sigma^0$ & ${2\over 3}-{a\over 3}(4\xi_1-3\epsilon)$& 
${2\over 3}-{a\over 3}(4\xi_1-3\epsilon)$ 
& $-{1\over 3}-{{2a}\over 3}(3\epsilon-\xi_2)$\\
$\Lambda^0$ & $-a\epsilon$&  $-a\epsilon$& 
$1-2a(\xi_2-\epsilon)$\\
$\Xi^0$ & $-{1\over 3}-{a\over 3}(-2\xi_1+5\epsilon+1)$& 
$-{a\over 3}(4\epsilon-1)$ & ${4\over 3}-{a\over 3}(8\xi_2-9\epsilon)$\\
\hline
   &$\sum\limits_q<L_z>_q^B$ & $\sum\limits_{\bar q}<L_z>_{\bar q}^B$&
$\sum\limits_q<L_z>_{q+\bar q}^B$\\
\hline 
p & $(2+\delta)\kappa a\xi_1$& $\kappa a\xi_1$&$a\xi_1$
\\
$\Sigma^+$ & $(2+\delta){{\kappa a}\over 3}(4\xi_1-\xi_2)$& 
${{\kappa a}\over 3}(4\xi_1-\xi_2)$& ${a\over 3}(4\xi_1-\xi_2)$
\\
$\Lambda^0$&$(2+\delta)\kappa a\xi_2$& $\kappa a\xi_2$&$a\xi_2$
\\
$\Xi^0$ & $(2+\delta){{\kappa a}\over 3}(4\xi_2-\xi_1)$& 
${{\kappa a}\over 3}(4\xi_2-\xi_1)$& ${a\over 3}(4\xi_2-\xi_1)$
\\
\hline
\end{tabular}
\end{center}
\end{table}

\begin{table}[ht]
\begin{center}
\caption{Quark spin and orbital angular momentum in other octet baryons.}
\begin{tabular}{cccccccccc} 
 Baryon   &    & $\Sigma^+$ &  &   & $\Lambda$ & & &$\Xi^0$& \\
&$\kappa=1/4$ & $\kappa=1/3$ &$\kappa=3/8$ &$\kappa=1/4$ &$\kappa=1/3$ 
& $\kappa=3/8$ & $\kappa=1/4$ & $\kappa=1/3$ &$\kappa=3/8$\\
\hline 
$<L_z>_u^B$  &0.130&0.141&0.147&0.038&0.028&0.023&0.014&0.000&$-$0.008\\ 
$<L_z>_d^B$  &0.096&0.071&0.058&0.038&0.028&0.023&0.023&0.017&0.014\\
$<L_z>_s^B$  &0.029&0.015&0.007&0.063&0.067&0.069&0.071&0.080&0.085\\
\hline 
$<L_z>_{\bar u}^B$&0.005&0.007&0.008&0.021&0.028&0.031&0.025&0.033&0.037\\
$<L_z>_{\bar d}^B$&0.053&0.071&0.079&0.021&0.028&0.031&0.013&0.017&0.019\\
$<L_z>_{\bar s}^B$&0.026&0.035&0.039&0.004&0.006&0.007&$-$0.001&$-$0.001& 
$-$0.002\\ 
\hline
$\sum\limits_q<L_z>_{q+\bar q}^B$ & 0.34&0.34&0.34 &0.18&0.18&0.18 &0.15& 
0.15&0.15  \\ 
\hline
$\Delta u^B$ & &0.84    & & & $-$0.07 & & & $-$0.29 &    \\ 
$\Delta d^B$ & &$-$0.17 & & & $-$0.07 & & & $-$0.04 &    \\ 
$\Delta s^B$ & &$-$0.35 & & &    0.77 & & &   1.05  &  \\ 
\hline
${1\over 2}\Delta\Sigma^B$ & & 0.16& & & 0.32& &  & 0.35&  \\ 
\hline
\end{tabular}
\end{center}
\end{table}

\begin{table}[ht]
\begin{center}
\caption{The quark spin and orbital content in the decuplet baryons.}
\begin{tabular}{cccc} 
 Baryon &$\Delta u^{B^*}$ & $\Delta d^{B^*}$& $\Delta s^{B^*}$
\\
\hline 
$\Delta^{++}$ & $3-3a(2\xi_1-\epsilon-1)$& $-3a$ & $-3a\epsilon$
\\
$\Delta^{+}$ & $2-a(4\xi_1-2\epsilon-1)$& $1-a(2\xi_1-\epsilon+1)$
&$-3a\epsilon$
\\
$\Sigma^{*0}$ & $1-2a\xi_1$& $1-2a\xi_1$&$1-2a\xi_2$
\\
$\Sigma^{*+}$ & $2-a(4\xi_1-\epsilon-2)$& $-a(\epsilon+2)$ &$1-2a\xi_2$
\\
$\Xi^{*0}$ & $1-a(2\xi_1+\epsilon-1)$& $-a(2\epsilon+1)$ 
&$2-a(4\xi_2-3\epsilon)$
\\
$\Omega^{-}$ & $-3a\epsilon$& $-3a\epsilon$ &
$3-6a(\xi_2-\epsilon)$
\\
\hline
&$\sum\limits_q<L_z>_q^{B^*}$ & $\sum\limits_{\bar q}<L_z>_{\bar
q}^{B^*}$& 
$\sum\limits_q<L_z>_{q+\bar q}^{B^*}$
\\
\hline 
$\Delta$&${{(2+\delta)\kappa a}}(3\xi_1)$&${{\kappa a}}(3\xi_1)$&$a(3\xi_1)$
\\
$\Sigma$&${{(2+\delta)\kappa
a}}(2\xi_1+\xi_2)$&${{\kappa a}}(2\xi_1+\xi_2)$&$a(2\xi_1+\xi_2)$
\\
$\Xi$& ${{(2+\delta)\kappa a}}(\xi_1+2\xi_2)$& ${{\kappa a}}(\xi_1+2\xi_2)$
&$a(\xi_1+2\xi_2)$
\\
$\Omega$&${{(2+\delta)\kappa a}}(3\xi_2)$& 
${{\kappa a}}(3\xi_2)$ &$a(3\xi_2)$
\\
\hline
\end{tabular}
\end{center}
\end{table}

\begin{table}[ht]
\begin{center}
\caption{Comparison of our predictions with data for the octet baryon
magnetic moments. The naive quark model (NQM) results are also listed. 
The quantity used as input is indicated by a star.} 
\begin{tabular}{cccccc} 
 Baryon    & data & & This paper  &  & NQM \\
     &           &$\kappa=1/4$&$\kappa$=1/3& $\kappa=3/8$& \\
\hline 
p     &     $ 2.79\pm 0.00$&  & 2.79$^{*}$ & &2.79$^{*}$\\
n     &     $-1.91\pm 0.00$&  &$-1.91^{*}$ & &$-1.91^{*}$\\
$\Sigma^+$ &$ 2.46\pm 0.01$&  & 2.67       & &  2.67\\
$\Sigma^-$ &$-1.16\pm 0.03$&  &$-1.09$     & &$-1.09$ \\
$\Lambda^0$&$-0.61\pm 0.00$&  &$-0.61^{*}$ & &$-0.61^{*}$ \\ 
$\Xi^0$    &$-1.25\pm 0.01$&  & $-1.43$    & &$-1.43$\\
$\Xi^-$    &$-0.65\pm 0.00$&  & $-0.49$    & &$-0.49$\\
\hline
$\mu_u$&  &2.404    &2.351    &2.328   &1.850 \\
$\mu_d$&  &$-1.047$ &$-0.944$ &$-0.892$&$-0.972$ \\
$\mu_s$&  &$-0.657$ &$-0.623$ &$-0.606$&$-0.613$ \\
\hline
\end{tabular}
\end{center}
\end{table}

\begin{table}[ht]
\begin{center}
\caption{Comparison of our predictions with data for the decuplet baryon 
magnetic moments. The naive quark model (NQM) results are also listed.} 
\begin{tabular}{cccc} 
 Baryon & data    & This paper  &  NQM \\
  &   & $\kappa=1/3$  &   \\
\hline 
$\Delta^{++}$ &$4.52\pm 0.50\pm 0.45$ \cite{bos91}&5.55&5.58 \\
 & $3.7~<~\mu_{\Delta^{++}}~<~7.5$ \cite{pdg00} &     &\\
$\Delta^{+}$  & $-$                   &  2.73  &  2.79 \\
$\Delta^{0}$  & $-$                   &$-0.09$ &  0.00 \\
$\Delta^{-}$  & $-$                   &$-2.91$ &$-2.79$\\ 
$\Sigma^{*+}$ & $-$                   &  3.09  &  3.11 \\
$\Sigma^{*0}$ & $-$                   &  0.27  &  0.32 \\
$\Sigma^{*-}$ & $-$                   &$-2.55$ &$-2.47$\\
$\Xi^{*0}$    & $-$                   &  0.63  &  0.64 \\
$\Xi^{*-}$    & $-$                   &$-2.19$ &$-2.15$\\
$\Omega^{-}$  &$-1.94\pm 0.17\pm 0.14$ \cite{diehl91}&$-1.83$&$-1.83$\\
 &$-2.024\pm 0.056$ \cite{wal95}& & \\
 &$-2.02\pm 0.05$ \cite{pdg00}  & & \\
\hline
$\mu_u$&  &2.351      &1.85 \\
$\mu_d$&  &$-0.944$   &$-0.97$ \\
$\mu_s$&  &$-0.623$   &$-0.61$ \\
\hline
\end{tabular}
\end{center}
\end{table}

\begin{figure}[h]
\epsfxsize=5.0in
\centerline{\epsfbox{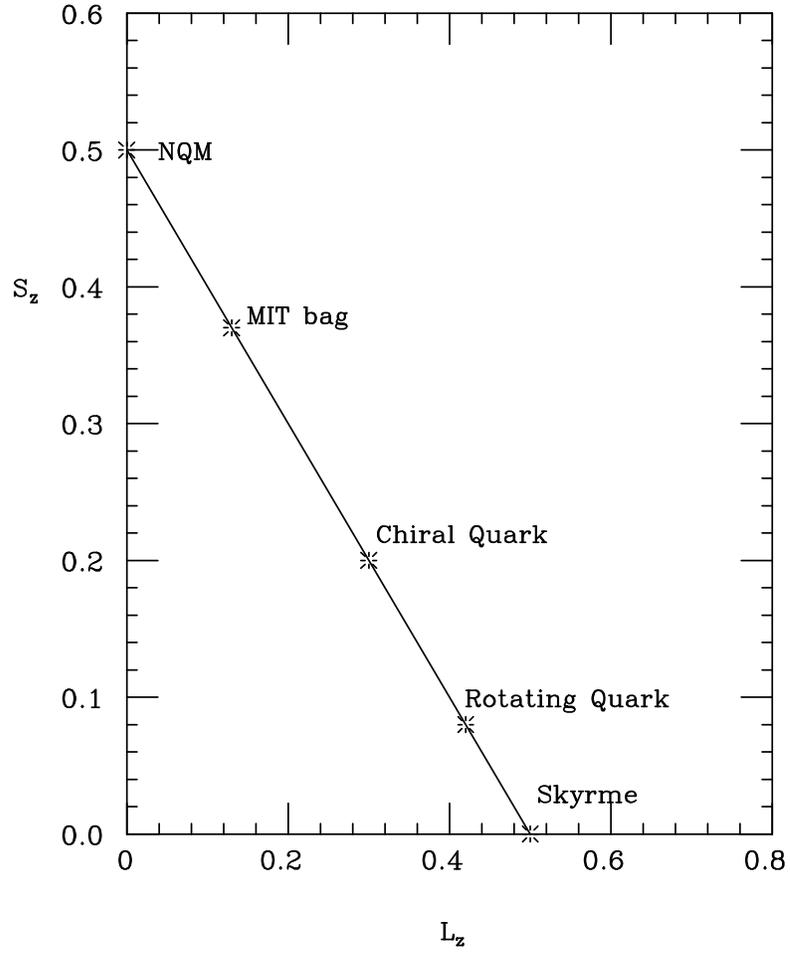}}
\caption{Quark spin and orbital angular momentum ($\sum\limits_q
<s_z>_{q+\bar q}$ versus $\sum\limits_q<L_z>_{q+\bar q}$) in the nucleon 
in different models.}
\end{figure}

\begin{figure}[h]
\epsfxsize=5.0in
\centerline{\epsfbox{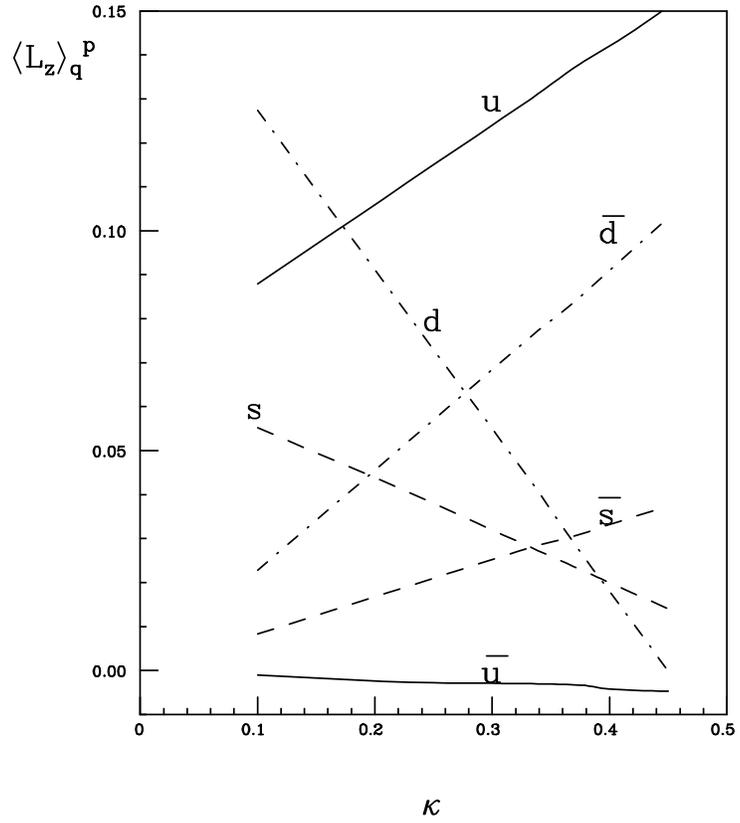}}
\caption{Quark or antiquark orbital angular momentum $<L_z>_{q,\bar q}$
in the proton as function of the partition factor $\kappa$.}
\end{figure}

\begin{figure}[h]
\epsfxsize=5.0in
\centerline{\epsfbox{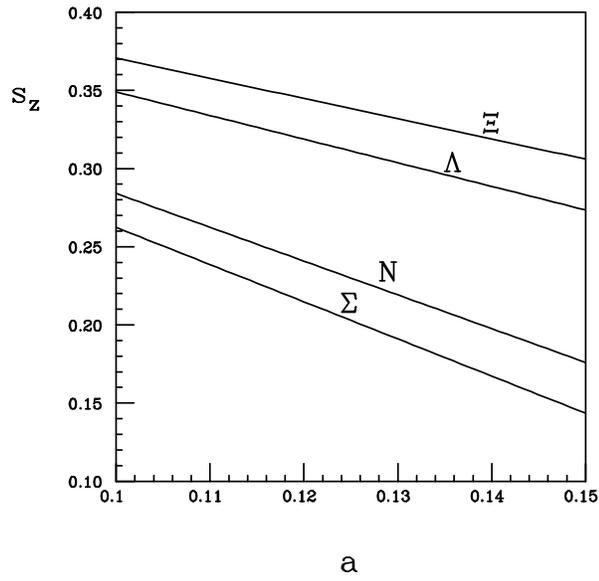}}
\caption{Quark spin content ($\sum\limits_q<s_z>_{q+\bar q}^B$) in
different octet
baryons as function of $a$}
\end{figure}

\end{document}